# TESTING OF ADVANCED TECHNIQUE FOR LINEAR LATTICE AND CLOSED ORBIT CORRECTION BY MODELING ITS APPLICATION FOR IOTA RING AT FERMILAB

A. Romanov[†], Fermilab, Batavia, IL

*Abstract*

Many modern and most future accelerators rely on precise configuration of lattice and trajectory. The Integrable Optics Test Accelerator (IOTA) at Fermilab that is coming to final stages of construction will be used to test advanced approaches of control over particles dynamics. Various experiments planned at IOTA require high flexibility of lattice configuration as well as high precision of lattice and closed orbit control. Dense element placement does not allow to have ideal configuration of diagnostics and correctors for all planned experiments. To overcome this limitations advanced method of lattice analysis is proposed that can also be beneficial for other machines. Developed algorithm is based on LOCO approach, extended with various sets of other experimental data, such as dispersion, BPM-to-BPM phase advances, beam shape information from synchrotron light monitors, responses of closed orbit bumps to variations of focusing elements and other. Extensive modeling of corrections for a big number of random seed errors is used to illustrate benefits from developed approach.

## INTRODUCTION

The Integrable Optics Test Accelerator (IOTA) is under construction at Fermilab. Its primary goal is to test advanced techniques for the stabilization of high intensity beams with highly nonlinear but integrable lattice designs. One set of experiments will be based on the use of a special nonlinear magnet that will create a big tune spread with two integrals of motion [1]. The second set of experiments will use an electron lens as a source of nonlinearity [2].

Simulations show that in order to benefit from nonlinear insertions, the linear lattice should be precisely tuned. Maximal errors for the main linear parameters of the lattice are listed in Table 1. Precise measurement and correction of the ring parameters will be done with beam-based model-dependent techniques. Figure 1 shows the IOTA ring with its main components.

Table 1. Maximal errors of the IOTA lattice for the integrable optics experiments

| Parameter | Max error |
|---|---|
| Betas at the insertion | 1 % |
| Beta beating | 3 % |
| Dispersion | 1 cm |
| Closed orbit at insertion | 0.05 mm |
| Phase advance between insertions | 0.001 |

aromanov@fnal.gov

To get the perfectly tuned linear lattice, IOTA will have a wide range of tools:
- Individual main quadrupole corrections
- Handy mechanical alignment design
- 20 combined X, Y and skew-quadrupole correctors
- 8 X-correctors in main dipoles
- 2 Y-correctors for the injection bump of the closed orbit
- 21 electrostatic pickups with precision of 1 μm
- Beam profile and position measurement monitors based on synchrotron light from main dipoles

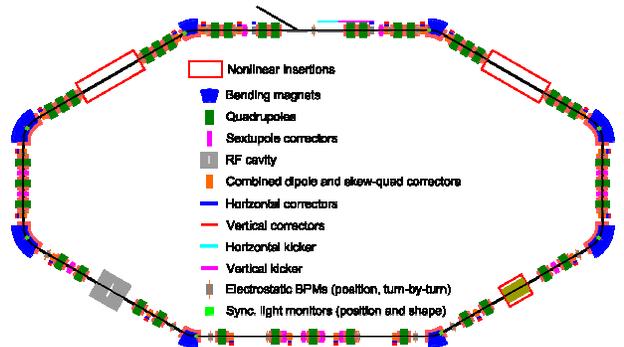

Figure 1. Schematic IOTA layout with its main components

## CORRECTION MODELING

Statistical analysis on a number of corrections applied to lattices with randomly introduced errors allows to study possible issues with the linear lattice and the closed orbit. "Sixdsimulation" software was used to analyze IOTA's configurations for all planned experiments with the nonlinear integrable systems.

The most important step in the study algorithm is the fit of pseudo-experimental data that can be done automatically or manually. In the manual mode, there are interactive tools for detailed analysis of the fit procedure. It is also possible to simulate several rounds of consecutive corrections, by using the corrected lattice at one iteration as the input for the next step.

*Fitting algorithm*

Both tasks of closed orbit and linear lattice corrections can be formulated as inverse problems when some set of experimental data $V_{exp,j}$ is available and the goal is to find the parameters $P_i$ of the model $M_j(P_i)$ that best describe the measurements. To find the approximate solution, the iterative method is used. The model parameters at iteration *(n)* are:

$$V_{\text{mod},j}^{(n)} = \mathbf{M}_j\left(P_i^{(n)}\right) \cdot s_j, \quad (1)$$

here $s_j$ is the normalization coefficients, that can be used to modify the weights of some experimental data points. In addition, both $V_{exp,j}$ and $V_{mod,j}$ are assumed to be normalized to the statistical error of $V_{exp,j}$.

The parameters of the model after iteration $(n)$ are:

$$P_i^{(n)} = P_i^{(0)} + \sum_{m=0}^{n-1} \Delta P_i^{(m)}, \quad (2)$$

The difference between the experimental data and the model is:

$$D_j^{(n)} = V_{\text{exp},j} - V_{\text{mod},j}^{(n)}, \quad (3)$$

The goal is to find such variation of the parameters $\Delta P_i^{(n)}$ that cancels residual difference between model and experimental data:

$$\Delta V_{\text{mod},j}^{(n)} = -\Delta D_j^{(n)} = D_j^{(n)}, \quad (4)$$

The model can be linearized in the case of small parameters variations:

$$\Delta V_{\text{mod},j}^{(n)} = s_j\left(\mathbf{M}_j(P_i^{(n)} + \Delta P_i^{(n)}) - \mathbf{M}_j(P_i^{(n)})\right) \approx$$
$$\approx s_j \left.\frac{\partial \mathbf{M}_j}{\partial P_i}\right|_{P_i^{(n)}} k_i \frac{\Delta P_i^{(n)}}{k_i} = M_{ij}^{(n)} \frac{\Delta P_i^{(n)}}{k_i}, \quad (5)$$

where $M_{ij}^{(n)}$ is the linearized and weighted model at iteration $(n)$:

$$M_{ij}^{(n)} = s_j k_i \left.\frac{\partial \mathbf{M}_j}{\partial P_i}\right|_{P_i^{(n)}}. \quad (6)$$

The model parameters variations can be obtained from here by applying pseudo inversion to the $M_{ij}^{(n)}$. Singular Value Decomposition (SVD) is a powerful method for such calculation. One of the remarkable features of this technique is easy control over the influence of the statistical error in the experimental data on the output result. Application of SVD gives the parameters variations at iteration $(n)$:

$$\Delta P_i^{(n)} = k_i \sum_j \left(M_{ij}^{(n)}\right)^{-1}_{\text{SVD}} D_j^{(n)}. \quad (7)$$

Summation over all iterations gives the total correction for the model parameters:

$$\Delta P_i = \sum_n k_i \sum_j \left(M_{ij}^{(n)}\right)^{-1}_{\text{SVD}} D_j^{(n)}. \quad (8)$$

In the case of no systematic errors, the $\chi^2$ function limit is:

$$\chi^2_{\min} = \sum_j D_j^{(n)} \to J - I, \quad (9)$$

where $J$ is the size of experimental data and $I$ is the number of fitting parameters.

## RESULTS

For orbit correction modelling, the misalignments of the quadrupoles and dipoles presented in Table 2 were used. All dipole correctors were involved in the closed orbit correction in all BPMs. Precision of the orbit measurements, which is expected to be on the order of 1 µm, was not accounted for because it is much smaller than the typical orbit error.

Table 2. Standard deviations of errors for closed orbit correction modeling

| Quads | Bends | |
|---|---|---|
| X, Y shifts | X,Y Shifts. | X,Y tilts |
| 0.1 mm | 0.1 mm | 0.06° |

Figure 2 illustrates maximal horizontal orbit error along the IOTA ring tuned for experiments with one nonlinear magnet for 1000 random errors before and after correction. The bottom two rows of Table 3 contain a summary of the orbit correction analysis for all four experiments. In each experiment, the orbit in the nonlinear insertion was corrected down to zero which means that it is limited only by the precision of BPMs.

In earlier work [2], it was found that synchronous rotation of a group of quadrupoles that creates axially symmetrical focusing and doesn't have BPMs or correctors in between is almost undetectable with a standard technique. As a result, the mean rotation angle of such a group remains uncorrected which may affect nonlinear dynamics.

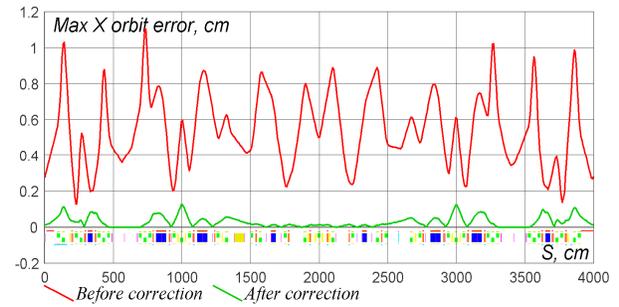

Figure 2. Horizontal orbit before and after correction for IOTA with one nonlinear magnet.

Table 3. Main lattice parameters before and after correction for nonlinear experiments at IOTA. "1NL" and "2NL" stands for one and two nonlinear insertions, "EL" and "MM" stands for electron and McMillan lenses.

| Parameter | 1NL Err | 1NL Fix | 2NL Err | 2NL Fix | EL Err | EL Fix | MM Err | MM Fix |
|---|---|---|---|---|---|---|---|---|
| $\langle|\Delta\nu_x|\rangle$ | 0.017 | $6.8\,10^{-5}$ | 0.012 | $5.3\,10^{-5}$ | 0.01 | $6.6\,10^{-6}$ | $9.5\,10^{-3}$ | $2.1\,10^{-5}$ |
| $\langle|\Delta\nu_y|\rangle$ | 0.018 | $1.0\,10^{-4}$ | 0.012 | $4.8\,10^{-5}$ | 0.009 | $6.9\,10^{-6}$ | $8.4\,10^{-3}$ | $1.5\,10^{-5}$ |
| Max[$\beta_{x,err}$], % | 41.3 | 0.12 | 45.1 | 0.046 | 29.2 | 0.025 | 62.3 | 0.06 |
| Max[$\beta_{y,err}$], % | 39.5 | 0.18 | 44.5 | 0.074 | 27.6 | 0.034 | 76.5 | 0.09 |
| Max[$D_{x,err}$], cm | 19.2 | 0.049 | 62.4 | 0.033 | 18.5 | 0.014 | 22.4 | 0.025 |
| Max[$D_{y,err}$], cm | 45.7 | 0.032 | 109 | 0.035 | 46.1 | 0.01 | 40.9 | 0.015 |
| RMS[$\alpha_{quad\,rot}$], deg | 0.95 | 0.027 | 0.95 | 0.044 | 0.96 | 0.007 | 0.98 | 0.012 |
| RMS[$G_{quad,err}$], % | 0.25 | 0.014 | 0.27 | 0.21 | 0.54 | 0.018 | 0.53 | 0.033 |
| $\langle$Max[$X_{err,ring}$]$\rangle$, mm | 4.5 | 0.54 | 9.8 | 0.54 | 11.9 | 0.46 | 4.5 | 0.37 |
| $\langle$Max[$Y_{err,ring}$]$\rangle$, mm | 2.7 | 0.34 | 6.1 | 0.33 | 3.0 | 0.31 | 2.1 | 0.29 |

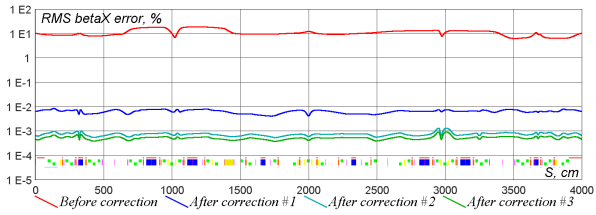

Figure 3. RMS errors of horizontal beta function before and after each of 3 consequent corrections for experiment with one nonlinear magnet.

In order to detect individual rotational errors of elements special set of experimental data was included, composed of responses of some predefined bumps to the focusing fields variations. Bumps should be created individually for every focusing element with maximal closed orbit distortion in it. Such data set equivalent to having weak corrector in every studied element, and thus allows to find its individual rotations.

The experimental data set for linear lattice correction modelling was composed of next values:
- Closed orbit responses to the dipole correctors measured with 1 µm precision.
- Betatron tunes with errors of $10^{-6}$.
- Dispersion measured with precision of 0.1mm.
- Responses of closed orbit bumps to the quadrupoles' variations

To study lattice correction 100 lattices were analysed for every experiment with randomly introduced errors described in Table 4. Each time lattice were corrected 3 consequent times. On average the first correction brings lattice parameters to conditions necessary for experiment, but for bigger errors it is necessary to do several iterations.

Figure 3 illustrates RMS error of horizontal beta function before and after corrections. Table 3 shows that there are no problems with correction of selected errors for all experiments with nonlinear integrable systems at IOTA.

## CONCLUSION

Closed orbit analysis shows that with proper alignment all planned experiments on nonlinear integrable systems at the IOTA should have orbit within physical aperture even without correction, which will greatly simplify first injection. After correction closed orbit inside nonlinear insertions will be corrected down to the precision of BPMs, which, without systematic errors will be within required margins. Study also shows that manual orbit scan is possible to independently align orbit tilt and position in both planes for all setups.

Analysis of linear lattice correction shows that with help of extended experimental data set it will be possible to correct essential parameters to match or exceed all requirements of the nonlinear experiments. There are several important limitation of presented analysis such as assumption of full linearity of the system and absence of unknown sources of errors, in other words there are no systematic errors. As it was mentioned, IOTA will have 8 beam profile monitors in electron's mode which can be used to cross check results of corrections. Systematic errors will also reveal itself as inability of fitting algorithm to minimize $\chi^2$ down to theoretical minimum.

Table 4. Standard deviations of errors for linear lattice correction modeling

| Quads | | BPMs | | Corr. calibr. |
|---|---|---|---|---|
| G | Rot. | Calibr. | Rot. | X&Y |
| 1 G/cm | 1° | 1 % | 1° | 2 % |